\documentclass[prb,twocolumn,aps,10pt]{revtex4-1}

\usepackage{graphicx}  
\usepackage{dcolumn}  
\usepackage{amssymb, amsmath}
\usepackage{lineno} 
\usepackage{braket}
\usepackage{bbold}
\def\mathbi#1{\textbf{\em #1}}
\begin{document}
\title{Numerical study of spin-dependent transition rates within pairs of dipolar and exchange coupled spins with (s=1/2) during magnetic resonant excitation}
\author{M.\ E.\ Limes}
\author{J.\ Wang }
\author{W.\ J.\ Baker}
\author{S.-Y.\ Lee}
\author{B.\ Saam}
\author{C.\ Boehme}
\email{boehme@physics.utah.edu}
\affiliation{Department of Physics and Astronomy, University of Utah, 115 South 1400 East, Salt Lake City, Utah, 84112-0830, USA}
\begin{abstract}
The effect of dipolar and exchange interactions within pairs of paramagnetic electronic states on Pauli-blockade-controlled spin-dependent transport and recombination rates during magnetic resonant spin excitation is studied numerically using the superoperator Liouville-space formalism.
The simulations reveal that spin-Rabi nutation induced by magnetic resonance can control transition rates which can be observed experimentally by pulsed electrically (pEDMR) and pulsed optically (pODMR) detected magnetic resonance spectroscopies.
When the dipolar coupling exceeds the difference of the pair partners' Zeeman energies, several nutation frequency components can be observed, the most pronounced at $\sqrt{2}\gamma B_1$ ($\gamma$ is the gyromagnetic ratio, $B_1$ is the excitation field).
Exchange coupling does not significantly affect this nutation component; however, it does strongly influence a low-frequency component $<\gamma B_1$.
Thus, pEDMR/pODMR allow the simultaneous identification of exchange and dipolar interaction strengths.
\end{abstract}
\maketitle
%
\section{Introduction}
In solids with weak spin-orbit coupling like silicon or carbon-based materials, spin-selection rules induced by spin conservation can drastically influence optical and electrical materials properties~\cite{Kaplan_1978,Lepine_1972,Frankevich_1992,Boehme_2003}. 
Because of this, a manipulation of spin states, e.g., by means of magnetic resonance, can change conductivity, luminescence, or absorption. 
These effects can be used for the investigation of the microscopic physical nature of the paramagnetic species involved in these processes, as is done with experimental techniques such as electrically (EDMR) and optically (ODMR) detected magnetic resonance spectroscopies.
An abundance of spin-dependent processes has been reported in the literature~\cite{Lepine_1972,Kaplan_1978,Frankevich_1992,McCamey_2008_2,Behrends_2010,Mueller_2000,Baker_2011}. 
Most of these reports involve the Pauli-blockade effect, where a transition of two paramagnetic states with $s=1/2$ into a single doubly occupied electron state with singlet configuration is controlled by the singlet content of the pair before the transition occurs. 
Such mechanisms are usually described by an ``intermediate-pair'' process, where an exclusive pair of two spins is formed.
This pair then either dissociates with spin-independent probability, or undergoes a transition into the singlet state that happens with probability $\propto|\langle S|\Psi\rangle|^2$, where $|\Psi\rangle$ is the spin state of the pair before the transition.
This intermediate-pair model, developed by Kaplan, Solomon, and Mott (KSM) in 1978~\cite{Kaplan_1978}, is distinct from other $s=1/2$ pair models which do not require the exclusive intermediate pair (see for instance the Lepine~\cite{Lepine_1972} model).
However, many experimental EDMR~\cite{Baker_2012,Morishita_2009} and ODMR~\cite{Lee_2011} studies have shown the validity of this picture for the description of several spin-dependent recombination and transport effects which involve transitions between localized electronic states. 
The KSM model is thus the basis for the calculation of spin-dependent transition rates presented here.

With the availability of high-power microwave sources, and the resulting development of pulsed electron paramagnetic resonance (EPR) techniques in the past 25 years, ODMR and EDMR have increasingly been conducted as transient, pulsed (p) experiments, on time scales where coherent spin-motion effects~\cite{Boehme_2003_2,McCamey_2008,Stegner_2006,Lee_2011} take place. 
Since coherent propagation of a quantum mechanical system is directly controlled by its Hamiltonian, this development in experimental techniques has dramatically enhanced access to the fundamental physical nature of the microscopic systems responsible for the EDMR and ODMR signals. 
Coherent spin effects such as spin echoes, spin--Rabi nutation or dynamic decoupling schemes have produced a variety of experimental insights. 
In order to fully utilize the spin effects observed with these techniques, a rigorous theoretical understanding of the signals is necessary. 
As pEPR spectroscopy evolved over the past decades, many studies have contributed to the development of this understanding~\cite{Schweiger_2001,Atherton_1993}, and it is now straightforward to derive information from pEPR data about Land\'e factors, spin-spin coupling phenomena, such as exchange coupling, dipolar coupling (which reveals distance between interacting spins), hyperfine couplings, and relaxation times, among other variables. 
Unfortunately, this rather comprehensive theory of pulsed EPR spectroscopy is only partially applicable to pulsed EDMR and ODMR experiments.
EDMR and ODMR are performed by measurement of spin-dependent rates whose observables depend on the permutation symmetry of the involved spin pairs, i.e., the singlet and triplet content.
 Thus, the observable of EDMR and ODMR experiments is fundamentally different than the observable of EPR experiments, the latter being the magnetic polarization of the observed spin ensemble. As a consequence, a spin ensemble that is simultaneously observed with EPR and EDMR/ODMR may exhibit entirely different signal behavior due to the different observables onto which the observed spin ensemble is projected.

Several recent studies aimed at developing and understanding pEDMR and pODMR signals have focused on electrically or optically detected transient nutation measurements, where a spectroscopy of observed spin--Rabi oscillation is conducted~\cite{Boehme_2003,Rajevac_2006,Gliesche_2008,Michel_2009,Glenn_2012,Glenn_2012_2, Lee_2012}. 
These studies have considered various spin--coupling regimes for the spin pair, including the absence of any spin-spin coupling~\cite{Boehme_2003,Rajevac_2006,Glenn_2012}, the presence of exchange interaction~\cite{Gliesche_2008}, and a disorder-induced distribution of spin-orbit interaction strengths~\cite{Michel_2009,Glenn_2012}.
Recently, the first analytical study of coherently controlled spin-dependent transition rates within pairs of strongly exchange- and dipolar-coupled pairs was conducted~\cite{Glenn_2012_2}.
However, a general numerical or analytical study for electrically or optically detected transient nutation of pairs with arbitrary spin-dipolar and spin-exchange interactions is lacking. 
Such a study is the focus of the present work.

\section{Intermediate-spin-Pair Model with dipolar and exchange coupling}
Following previous discussions of spin-dependent transition controlled by intermediate pairs~\cite{Kaplan_1978, Haberkorn_1980, Boehme_2003, Rajevac_2006, Gliesche_2008, Michel_2009, Glenn_2012,Glenn_2012_2}, we describe the dipolar- and exchange-coupled intermediate-spin-1/2 pairs with the Hamiltonian
\begin{equation} 
\begin{split}
\hat{H}_{\text{spin}} =  \hbar \bigl[ \mathbi{B} \cdot(\gamma_a \hat{\mathbi{S}}_a \! + \! \gamma_b \hat{\mathbi{S}}_b) &-  J \hat{\mathbi{S}}_a \cdot \hat{\mathbi{S}}_b  \\
&-D(3\hat{S}_a^z \hat{S}_b^z \!  - \! \hat{\mathbi{S}}_a \cdot \hat{\mathbi{S}}_b)\bigr].
\end{split}
\end{equation}
Here, the first term represents the Zeeman interaction for both spin-pair partners, the second term is an isotropic exchange interaction, the third is a secular (high-field approximation) magnetic-dipole coupling, $\gamma_a$ and $\gamma_b$ are the effective gyromagnetic ratios of the spin-pair partners $a$ and $b$, respectively.
The magnetic field
\begin{equation} 
\mathbi{B} = \hat{\mathbi{z}}B_0 \! + \! \hat{\mathbi{x}}B_1e^{-i\omega t}
\end{equation}
consists of a static component $B_0$ along the $\hat{\mathbi{z}}$-axis, and an oscillating component that is chosen to be along the $\hat{\mathbi{x}}$-axis.
When the exchange interaction strength $J$ and the dipolar coupling strength $D$ are scaled by $\hbar$, they can be directly compared to the Larmor separation $\Delta \omega$ of the electron and hole. 
The negative signs in front of the $D$ and $J$ terms are chosen to represent an attractive electron-hole pair~\cite{Abragam_1970, Anderson_1963, Elliott_1963}. 
We note that changing the sign of sign of $J$ and/or $D$ will not change the results presented below (such a sign change could occur for like-charge spin pairs, e.g., bipolarons~\cite{Behrends_2010}).

The spin-pair Hamiltonian in absence of radiative excitation ($B_1 \! = \! 0$) is rotated into an energy eigenbasis by a Jacobi rotation, $\hat{H}_{\text{en}} \! = \! U^{\dagger}\hat{H}_{\text{spin}}U$, with the resulting eigenbasis given by
\begin{equation}
\label{basis}
U^{\dagger} \left( \begin{array} {c}
			\ket{\uparrow \uparrow} \\
			\ket{\uparrow \downarrow} \\
			\ket{\downarrow \uparrow}  \\
			\ket{\downarrow \downarrow} \\
			\end{array}			\right) =
			\left( \begin{array} {c}
			\ket{\uparrow \uparrow} \\
  			\cos(\phi)\ket{\uparrow \downarrow} \! - \! \sin(\phi)\ket{\downarrow \uparrow} \\
			\cos(\phi)\ket{\uparrow \downarrow} \! + \! \sin(\phi)\ket{\downarrow \uparrow} \\
			\ket{\downarrow \downarrow} \\
			\end{array} \right)~,
\end{equation} where $\cot(2\phi) = \frac{\Delta \omega}{J  -  D}$.
In the case of either strong dipolar or strong interaction exchange, the energy eigenbasis becomes a set of singlet and triplet states.
With strong dipolar coupling, $\phi \! \rightarrow \! -\frac{\pi}{4}$ ($D \! \rightarrow \! \infty, ~J \! \rightarrow \! 0$), the energy eigenbasis becomes $\lbrace \ket{T_+},\ket{T_0},\ket{S},\ket{T_-} \rbrace$; strong exchange coupling, $\phi \! \rightarrow \! \frac{\pi}{4}$ ($J \! \rightarrow \! \infty,~D \! \rightarrow  \! 0$ ), produces an energy eigenbasis $\lbrace \ket{T_+},\ket{S},\ket{T_0},\ket{T_-} \rbrace$.
In either one of these strong coupling cases, the only ESR-allowed transitions are those within the triplet manifold, leading to a strong triplet ESR signal.
However, because the triplet-singlet transition probability is zero, there is no observable pODMR/pEDMR signal.
Any intermediate case (e.g., $J \approx D \approx \Delta \omega$) will have an energy eigenbasis of $\lbrace \ket{T_+},\ket{2}, \ket{3}, \ket{T_-} \rbrace$, where $\ket{2}$ and $\ket{3}$ will each have a mixture of singlet and triplet content defined by the relative magnitudes of the dipolar and exchange strengths. 
Therefore, the transitions between states are uniquely governed by the collection of system parameters  $D$, $J$, and $\Delta\omega$.

For pODMR/pEDMR experiments on intermediate-spin-pair processes, the observable depends on the permutation symmetry of the individual pairs, contrary to most conventional spectroscopy experiments which probe polarization states.
An extended discussion of such intermediate-pair related pEDMR/pODMR observables is given by Gliesche et al.~\cite{Gliesche_2008}, who established the connection of the spin density operator $\hat\rho$ of an ensemble of spin-$\frac{1}{2}$ pairs to a spin-dependent rate transient 
\begin{equation}
\label{observable}
Q(\tau)= \int_0^{t_0}R(t)dt = \displaystyle\sum\limits_{i=i}^4 (\hat{\rho}_{ii}(\tau) \! - \! \hat{\rho}_{ii}^S)(1 \! - \! e^{-r_it_0}),
\end{equation}
which follows coherent spin excitation.
In Eq.~\ref{observable}, the density matrix is in the $4\times4$ energy eigenbasis representation and the time-dependent function $R(t)$ is the spin-dependent rate after the pulse excitation, which is assumed to end at $t=0$.
Since $R(t)$ is a current for pEDMR experiments, the integral $Q(\tau)$ becomes a number of charge carriers which undergo spin-dependent transitions due to the resonant spin excitation.
The dependence of $Q$ on the pulse length $\tau$ will reveal information about how the density operator $\hat\rho$ evolves from the steady state to a coherent state due to the resonant excitation.
Thus, $Q(\tau)$ is an easily accessible observable for the coherently manipulated spin ensemble, representing either the number of charge carriers (for pEDMR) or photons (for pODMR).

The transient evolution of $Q(\tau)$ during the pulse can be Fourier transformed (FFT$\left\lbrace Q(\tau) \right\rbrace$) in order to make the frequency components of the coherent spin motion explicit.
A comparison of experimentally obtained Rabi frequency spectra with calculations we present below gives insight into the nature of the spin-pair Hamiltonian.
As the spin-pair Hamiltonian crucially depends on the microscopic nature of the spin pairs, pEDMR/pODMR experiments are superb probes to gain unambiguous experimental access to spin-dependent transport and recombination processes.

Again following previous descriptions~\cite{Haberkorn_1980,Boehme_2003,Rajevac_2006, Gliesche_2008, Michel_2009, Glenn_2012,Glenn_2012_2}, we describe the evolution of the density operator $\hat\rho$ by a stochastic Liouville equation
\begin{equation}
\label{nonrot}
\partial_t\hat{\rho} = \frac{i}{\hbar} [\hat{\rho},\hat{H}_{\text{en}}] + S[\hat{\rho}],
\end{equation}
where the stochastic term $S[\hat{\rho}] \! = \!S_{\text{cr}}[\hat{\rho}] \!+\! S_{\text{an}}[\hat{\rho}]$ is the sum of creation and annihilation terms of the spin pairs.

As shown elsewhere~\cite{Boehme_2003}, the recombination probabilities for the different energy eigenbasis states are given by $r_i  = r_S \left|\braket{i|S}\right |^2 + r_T \left|\braket{i|T}\right|^2$, where $r_S$ and $r_T$ are the singlet and triplet recombination probabilities, respectively.
Using Eq.~\ref{basis}, the various recombination rates can be expressed in terms of the coupling parameters by
\begin{eqnarray}
\label{rates}
r_{1,4} &=& r_{T}\\ \nonumber
r_{ 2,3 } &=& \frac{ 1 }{ 2 } r_{ S }\left( 1 \! \mp \! \frac { J \! - \! D }{ \sqrt { (J \! - \! D)^{ 2 } \! + \! \hbar \Delta \omega )^{ 2 } }}  \right)\\
&& + \frac { 1 }{ 2 } r_{ T }\left(1 \! \pm \! \frac{J \! - \! D}{\sqrt{(J \! - \! D)^{2} \! \! + \! (\hbar\Delta\omega)^2}} \right).
\end{eqnarray}
The eigenstates $\ket{1}$ and $\ket{4}$ always remain pure triplet states ($|T+\rangle$ and $|T-\rangle$, respectively); their recombination rates are thus not affected by any coupling within the spin pair.
Under strong coupling such as $D\gg \!\Delta\omega$, $r_2 \! = \! r_T $ and $r_3 \! = \! r_S$ ($J\gg \!\Delta\omega$, $r_2 \! = \! r_s $ and $r_3 \! = \! r_T$).
The dissociation rate coefficient $d$ is assumed to be spin independent. In the energy eigenbasis, the stochastic annihilation term $S_{\text{an}}[\hat{\rho}]$ has matrix elements in a convenient form, $\left\lbrace S_{\text{an}}[\hat{\rho}] \right\rbrace_{ij} = (r_i + r_j + 2d)\frac{\rho_{ij}}{2}$.
We also assume pair generation only creates pairs in an energy eigenstate, $\left\lbrace S_{\text{cr}}[\hat{\rho}] \right\rbrace_{ij} = \delta_{ij}\frac{k}{4}$, where $k$ is the net generation rate of all four states.
This creation term is the only inhomogeneous contribution to Eq.~\ref{nonrot}.
In this paper we neglect the Redfield relaxation matrix, an assumption that is valid in the short-time regime ($\tau < \frac{1}{r_{S}}\sim T_{2} < \frac{1}{r_{T}} \ll T_{1}$).
For the purpose of obtaining sufficient resolution, some pulse lengths violated this assumption (see Fig.~\ref{fig:dip_only}(d) and (g)).

\section{Analytical and Numerical methods}
\label{anal}
In the following section we outline our study of the observable $Q(\tau)$ that results from the coherent excitation of the spin pair.
Eq.~\ref{nonrot} is a set of sixteen coupled inhomogeneous ordinary differential equations (ODEs) that were previously solved using a Runge-Kutta or comparable ODE solver~\cite{Rajevac_2006,Gliesche_2008,Michel_2009}.
These computationally intensive methods make the convolution of distributions of many parameters ($J$, $D$, bandwidth of pulse, etc.) impractical without a supercomputer. 
We use two techniques that lead to a significant decrease in computation time.
In Sect.~\ref{rotfram} we detail the first step of the computation, a transformation into the rotating frame.

Once in the rotating frame, several limiting cases of the Rabi nutation frequencies are demonstrated in Sect.~\ref{Rabilim}.
The limiting cases of overall weak coupling, strong exchange coupling, strong dipolar coupling, and a large difference in dipolar and exchange coupling are described.
Sect.~\ref{rotfram} also includes an analytical description of the $\sqrt{2}\gamma B_{1}$ Rabi frequency component that occurs in the presence of strong dipolar coupling.
These limiting cases provide significant insight into qualitative features observed in the numerically calculated general gases, such as resonance location, Rabi frequency, and signal amplitude.

In addition to the use of the rotating frame, the calculation of the time-dependent change of the density matrix was aided by the use of Liouville-space formalism, and is discussed in Sect.~\ref{Liouvillespace}.
A direct consequence of this formalism is that the inhomogeneous stochastic Liouville equation is cast into a readily tractable and solvable form.
Compared to previous work~\cite{Rajevac_2006,Gliesche_2008}, the speed of the simulation allows us to perform a larger and more detailed study of $Q(\tau)$'s dependence on dipolar $D$ and exchange $J$ interactions with respect to the Larmor frequency separation $\Delta\omega$ and the excitation-field strength $B_{1}$. 

In Sect.~\ref{reslut}, representative results of these simulations are given and discussed. Using the methods from Sect.~\ref{anal}, we simulate $Q(\tau)$ for a range of values $D$ and $\Delta\omega$ with a fixed excitation field $B_{1}$.
Then, $Q(\tau)$ is simulated as a function of $D$ and $J$ with a fixed $\Delta\omega$ and $B_{1}$.
Finally, $Q(\tau)$ is simulated with small and large exchange coupling strengths, along with a complete Pake distribution of dipolar interaction strengths.

\subsection{Rotating-Frame Stochastic Liouville Equation}
\label{rotfram}
The rotating frame corresponds to a transformation of the Hamiltonian from the energy eigenbasis: $\hat{H}_R = R^{\dagger} \hat{H}_{\text{en}} R$.
The rotating-frame density matrix is then given by $\hat{\rho}_R = R^{\dagger} \hat{\rho} R$.
Here $R = R^{\frac{1}{2}}_z \! \otimes \! R^{\frac{1}{2}}_z$ is the $4\! \times \! 4$ spin-$\frac{1}{2}$ pair rotation operator, and $R^{\frac{1}{2}}_z$ is the rotation operator for a spin-$\frac{1}{2}$ state around the $z$-axis by an angle $\omega t$.
The resulting rotating-frame Hamiltonian is
\begin{widetext}
 \begin{equation}
 \label{H_R}
 \hat{H}_R =\frac{\hbar}{2} \left( \begin{array}{cccc}
  2\omega_0 \! - \! \frac{J}{2} \! - \! D & \gamma B_1 (\cos(\phi)\!  -\!  \sin(\phi)) & \gamma B_1 (\cos(\phi)\!  + \! \sin(\phi)) & 0\\
 \gamma B_1 (\cos(\phi)\!  -\!  \sin(\phi)) & \frac{J}{2}\!  +\!  D\!  +\!  \sqrt{\Delta\omega^2 \!  + \! (J \! - \! D)^2} & 0 & \gamma B_1 (\cos(\phi) \!  - \!  \sin(\phi))\\
  \gamma B_1 (\cos(\phi) \!  + \!  \sin(\phi)) &  0&   \frac{J}{2} \!  + \! D \! -\!  \sqrt{\Delta\omega^2 \! +\!  (J \! - \! D)^2} & \gamma B_1 (\cos(\phi) \! + \!  \sin(\phi))\\
    0& \gamma B_1 (\cos(\phi) \!  - \!  \sin(\phi))  & \gamma B_1 (\cos(\phi) \! +\!  \sin(\phi))& -2\omega_0 \! - \!  \frac{J}{2} \!  - \! D\\
 \end{array}\right)
 \end{equation}
\end{widetext}
and has no explicit time dependence.
Note that the energy levels for the energy eigenbasis $E_1$, $E_2$, $E_3$, and $E_4$ reside on the diagonal.
We label the average of the spin-pair Larmor frequencies $\omega_0 \! = \! (\omega_a \! + \! \omega_b)/2$ and Larmor frequency separation  $\Delta \omega \! = \! \omega_a \! - \! \omega_b$.
We assume that the Rabi frequency of each spin is the same ($\gamma_a B_1\! \approx \! \gamma_b B_1$), allowing us to explicate the results in terms of a single on-resonance Rabi frequency $\gamma B_1$.
Neglecting the very small difference in the individual-spin Rabi frequencies symmetrizes the simulations about $\omega \! - \! \omega_0 \! = \! 0$, rather than demonstrating an inconsequential asymmetry.
After an additional time-independent term $\hat{F} = R^{\dagger}\partial_t R$ is absorbed into an effective Hamiltonian $\hat{H} = \hat{H}_R \! - \! \hat{F}$, the rotating-frame stochastic Liouville equation becomes
\begin{equation}
\label{stochL}
\partial_t\hat{\rho}_R = \frac{i}{\hbar} [\hat{\rho}_R,\hat{H}] + S[\hat{\rho}_R].
\end{equation}
As expected from this transformation, the only term left with time dependence in Eq.~\ref{stochL} is the rotating-frame density matrix $\hat{\rho}_R$.

\subsection{Limiting Cases of the Rabi Frequencies}
\label{Rabilim}
Useful equations that elucidate limiting cases can be derived from finding the single-transition Rabi frequencies of the rotating-frame Hamiltonian given in Eq.~\ref{H_R}.
By considering an induced transition between only two of the available four states and solving a $2 \! \times \! 2$ eigenvalue problem, it can be shown that the single-transition Rabi frequencies are
\begin{equation}
\label{Rabio}
\Omega_{ij} = \sqrt{(1 \! \mp \! \sin 2\phi) \left(\gamma B_1\right)^2 \! + \! (\omega \! - \! \omega_{ij})^2}~\!.
\end{equation}
The negative sign in the first term under the radical on the right hand side gives the Rabi frequencies for the $\ket{T_{\pm}} \! \leftrightarrow \! \ket{2}$ transitions between the pure triplet states and $\ket{2}$ state  ($(i,j) = \lbrace(T_{+},2);(T_{-},2) \rbrace$).
The plus sign in Eq.~\ref{Rabio} gives the Rabi frequencies for the $\ket{T_{\pm}} \!  \leftrightarrow \! \ket{3}$ transitions between the pure triplet states and $\ket{3}$ energy eigenstate ($(i,j) = \lbrace(T_{+},3);(T_{-},3) \rbrace$).
In general, there are four resonant frequencies, $\omega_{ij} =(E_i - E_j)/\hbar$.
If an on-resonant excitation frequency $\omega$ is applied such that $\omega = \omega_{ij}$, the second term in Eq.~\ref{Rabio} vanishes.

\subsubsection{Weak and Effectively Weak Coupling}
\label{weak}
For the first limiting case, let the coupling terms $J$ and $D$ approach zero. 
In this weak-coupling regime, $|J| \! + \! |D| \! \ll \! \Delta\omega$, the first term in Eq.~\ref{Rabio} tends towards the limit $(1 \! \mp \! \sin 2\phi)\!  \rightarrow \! 1$ and there is only an on-resonance Rabi oscillation frequency of a single uncoupled spin, $\gamma B_1$. 
There are two resonant transitions with a two-fold degeneracy corresponding to the Larmor frequency of each spin in the pair.
If the Larmor separation $\Delta\omega$ is zero (indicating that the gyromagnetic ratios of the electron and the hole are the same), there is only one transition that has a degeneracy of four and a Rabi frequency $\gamma B_1$. 
However, if there is a sufficient excitation-field strength $B_{1}$, both uncoupled spins will nutate coherently, creating a spin-beating effect with a 2$\gamma B_1$ Rabi frequency component~\cite{McCamey_2010, Baker_2011}.

For the second limiting case consider an effectively weak coupling, where the difference in coupling strengths becomes much less than the Larmor separation, $|J \! - \! D| \! \ll \! \Delta\omega$. In this limit there are four non-degenerate resonant transitions.
As in the weak regime, a pair in the effectively weak regime has a Rabi frequency $\gamma B_1$ equal to that of a single uncoupled spin.

Both weak and effectively weak coupling leave the energy eigenbasis completely unaffected by the rotation performed in Eq.~\ref{basis}.
In the latter case, this happens even though the couplings $J$ and $D$ could individually be quite large compared to $\Delta\omega$.
However, the resonance frequencies for each transition will be shifted due to the increased coupling strengths.
This non-degenerate energy spectrum distinguishes the effectively weak coupling from weak coupling; see Fig.~\ref{fig:ex_and_dip}.
\subsubsection{Strong Dipolar Coupling}
\label{dipole}
Now consider the limiting case of strong dipolar coupling, $|D| \gg | \Delta \omega |$,  with no exchange coupling, $J = 0$.
As $D$ gets large, $\sin 2\phi \! \rightarrow \! -1$, and the four resonant single-transition frequencies, offset from $\omega_0$, are approximately
\begin{equation}
\label{strD}
 \omega_{\pm,2} \approx \pm(\frac{3D}{2} \! + \! \frac{\Delta \omega^2}{4D}),~\omega_{\pm,3} \approx \pm(\frac{D}{2} \! - \! \frac{\Delta \omega^2}{4D}).
\end{equation}
The first term under the radical on the right hand side of Eq.~\ref{Rabio} is $(1-\sin 2\phi) \! \rightarrow \! 2$ for the $\ket{T_{\pm}} \!  \leftrightarrow \! \ket{2}$ transitions, and $(1+\sin 2\phi)\!  \rightarrow\!  0$ for the $\ket{T_{\pm}} \! \leftrightarrow \! \ket{3}$ transitions.
Therefore, strong dipolar coupling within the pair yields an on-resonance Rabi frequency of $\sqrt{2}\gamma B_1$ for each transition between the pure triplet states and the $\ket{2}$ state.
The $T_{\pm} \! \leftrightarrow \! \ket{2}$ transition probabilities are large but have an overall reduction of the pEDMR/pODMR signal, owing to the strong triplet character of the $\ket{2}$ state.

We predict a $\sqrt{2}\gamma B_1$ Rabi frequency for any spin-$\frac{1}{2}$ pair with sufficient Larmor separation and strong enough dipolar coupling.
When strongly coupled, an applied excitation necessarily affects both spins in a pair, even if only a monochromatic excitation is applied.
The strong dipolar coupling (like a strong exchange coupling~\cite{Gliesche_2008}) allows access to only one quantum state, and prohibits isolating an individual spin within the spin pair.
This behavior has been well known from traditional magnetic resonance spectroscopy~\cite{Weber_2002,Schweiger_2001} and, without explicit theoretical proof, it has already been applied to experimental pODMR~\cite{Lips_2005,Herring_2009} and pEDMR~\cite{Herring_2009,Lee_2010} data.

\subsubsection{Strong Exchange Coupling}
\label{exchange}
We now consider the strong exchange coupling regime, where $|J| \gg | \Delta \omega |$, with no dipolar coupling, $D = 0$.
As $J$ gets large, $\sin 2\phi \rightarrow 1$, and the resonant single-transition frequencies, offset from $\omega_0$, are approximately
\begin{equation} 
 \omega_{\pm, 2} \approx \pm\left(J + \frac{\Delta \omega^2}{4J}\right)~,~\omega_{\pm, 3} \approx \mp \frac{\Delta \omega^2}{4J}.
 \label{exchange_pos}
\end{equation}
The first term under the radical on the right hand side of Eq.~\ref{Rabio} is $(1 \! + \! \sin 2\phi) \! \rightarrow \! 2$ for the $\ket{T_{\pm}} \!  \leftrightarrow \! \ket{3}$ transitions, and $(1 \! - \!\sin 2\phi)\!  \rightarrow\!  0$ for the $\ket{T_{\pm}} \! \leftrightarrow \! \ket{2}$ transitions.
The single-transition analysis predicts a $\sqrt{2}\gamma B_1$ Rabi frequency for the $\ket{T_{\pm}} \!  \leftrightarrow \! \ket{3}$ transitions.
However, this naive analysis does not take into account that the splitting in $\ket{T_{\pm}} \!  \leftrightarrow \! \ket{3}$ transition frequencies is so small that the transitions will be driven simultaneously by $B_1$.
Therefore, Eq.~\ref{Rabio} is no longer valid, and a multiple-transition analysis must be used.
The $\ket{T_{\pm}} \!  \leftrightarrow \! \ket{2}$ transitions are far away from $\omega_0$ and have Rabi frequencies approaching zero.
Because of this, the strong exchange-coupling regime can be analyzed using only the $\ket{T_{\pm}} \! \leftrightarrow \! \ket{3}$ transitions.
If a excitation frequency of $\omega = \omega_0$ is applied, two of the three rotating-frame energy eigenvalues in the multiple-transition analysis are degenerate.
This simplifies the eigenvalue problem significantly, and the (three-state) Rabi frequency is found to be
\begin{equation}
\Omega = \sqrt{\frac{\Delta \omega^2}{4J} + 2(1 \! + \! \sin 2 \phi )(\gamma B_1)^2} \approx 2 \gamma B_1~,  \end{equation}
for the $\ket{T_{+}} \! \leftrightarrow \! \ket{3} \! \leftrightarrow \! \ket{T_{-}}$ transition.
We note that the single-transition Rabi frequencies for $\ket{T_{\pm}} \! \leftrightarrow \! \ket{3}$ do not merely add to a total $2 \sqrt{2} \gamma B_1$ Rabi frequency.

 If the power of the excitation field is lowered, only a $\gamma B_1$ Rabi frequency is seen in the case of uncoupled pairs, whereas a strongly exchange-coupled pair always has a $2\gamma B_1$ component, provided the signal is strong enough.
 This fact has served to distinguish uncoupled and strongly exchange-coupled states in experimental studies\cite{McCamey_2010, Baker_2011}.

\subsubsection{Large Difference in Exchange and Dipolar Strengths}
\label{both}
The final limiting case we consider is to take the difference in coupling strengths to be large with respect to the separation of the Larmor frequencies, and the exchange strength to be greater than the dipolar strength, $J \! - \! D \! \gg \! \Delta \omega$. 
In this limit we have $(1 \! - \! \sin 2\phi) \! \rightarrow \! 0$ for the $\ket{T_{\pm}} \! \leftrightarrow \! \ket{2}$ transitions, and $(1 \! + \! \sin 2\phi) \! \rightarrow \! 2$ for the $\ket{T_{\pm}} \! \leftrightarrow \! \ket{3}$ transitions.
The resonant single-transition frequencies, offset from $\omega_0$, are now
\begin{equation}
\label{diff}
\begin{array}{c}
  \omega_{\pm, 2} \approx \pm\left(J \! + \! \frac{D}{2} \! + \! \frac{\Delta \omega^2}{4(J \! - \! D)}\right)~, \\
  \omega_{\pm, 3} \approx \pm \left(\frac{3D}{2} \! - \! \frac{\Delta \omega^2}{4(J \! - \! D)}\right).
 \end{array}
\end{equation} 
The presence of dipolar coupling splits the transition frequencies enough that the single-transition analysis for Eq.~\ref{H_R} becomes valid again.
Therefore, in the limit of a large difference in dipolar and exchange coupling strengths, a Rabi frequency of $\sqrt{2}\gamma B_1$ will occur when on resonance with the $\ket{T_{\pm}} \! \leftrightarrow \! \ket{3}$ transitions, and the $\ket{T_{\pm}} \! \leftrightarrow \! \ket{2}$ transitions have a vanishingly small transition probability.

The other limit in the strong-coupling regime that we do not describe in detail is the difference in coupling strengths large with respect to the separation in Larmor frequencies, and the dipolar strength is greater than the exchange strength, $D \! - \! J \! \gg \! \Delta \omega$.
An analysis similar to that given above shows that the $\sqrt{2} \gamma B_1$ Rabi frequency components exist, but have resonances far away from the central average $\omega_0$ of the spin-pair Larmor frequencies.

We will refer to these limiting cases as we discuss the features appearing in the results of the following simulations.

\subsection{Liouville-Space Formalism}
\label{Liouvillespace}
We now reformulate the rotating-frame description using Liouville operator space, also known as superoperator formalism~\cite{Ernst_1987,Happer_2010}, to increase the computational power of the simulation.
This technique was also used recently in a model for magnetic-field effects in disordered semiconductors~\cite{Shushin_2011}.
The essence of this reformulation is the representation of the state population as a $16 \! \times \! 1$ column vector $\rho$ instead of the typical $4 \! \times \! 4$ density matrix $\hat{\rho}$. 
Operations involving $\hat{A}$ are associated with corresponding superoperators $A$.
Note that this formalism produces no new physics, but simply recasts the problem such that a convenient, tractable solution to Eq.~\ref{stochL} is obtained.

Using superoperator formalism, the rotating-frame inhomogeneous stochastic Liouville equation (Eq.~\ref{stochL}) can be rewritten in the compact form
\begin{equation}
\label{rat}
\partial_t\rho_{R} \! = \! \frac{i}{\hbar} H  \rho_R \! + \! S_{\text{an}}  \rho_R + K \! = \! G   \rho_R \! + \! K.
\end{equation}
Here $H \! \rho_R$ is the abbreviated superoperator form of the commutator $[\hat{\rho}_R, \hat{H}]$. $H$ is a $16 \times 16$ superoperator that can be written as
\begin{equation} 
H \!= \! \left( \begin{array}{cccc}
	\hat{H} \! - \! IH_{11} & IH_{12} & IH_{13} & IH_{14}\\
	IH_{21} & \hat{H} \! - \! IH_{22} & IH_{23} & IH_{24}\\
	IH_{31} & IH_{32} & \hat{H} \! - \! IH_{33} & IH_{34}\\
	IH_{41} & IH_{42} & IH_{43} & \hat{H} \! - \! IH_{44}\\
\end{array}\right),
\end{equation}
where $I$ is the $4\times4$ identity matrix and $H_{ij}$ are the matrix elements of the $4 \times 4$ Hamiltonian $\hat{H}$.
In Eq.~\ref{rat}, $S_{\text{an}} $ is a time-independent diagonal $16\times 16 $ matrix of the appropriate stochastic annihilation terms corresponding to $S_{\text{an}}[\hat{\rho}]$.
The creation term $K$ is a time-independent $16\times 16 $ matrix consisting of the appropriate stochastic creation/generation terms corresponding to $S_{\text{cr}}[\hat{\rho}]$ and is the sole inhomogeneous part of Eq.~\ref{rat}.
 The superoperator $G$ is merely the addition of $\frac{i}{\hbar}H$ and $S_{\text{an}}$; it is a symmetric and relatively sparse matrix with 160 zeroes.

A steady-state density matrix $\rho^S$ is used to define $\rho(0)$, the density matrix at time $t \! = \! 0$, and is obtained by neglecting the coherent excitation ($B_1 \! = \! 0$) and finding a steady-state superoperator $G_S$ from Eq.~\ref{rat}.
Using the variation-of-parameters method, the ODE in Eq.~\ref{rat} is solved analytically by
\begin{eqnarray}
\label{sol}
 \rho_R(t) = e^{Gt}(\rho(0) \! + \! G^{-1}K) \! - \! G^{-1}K,\\ \nonumber
  \rho(0) = G^{-1}_S K,
\end{eqnarray}
where $\rho(0)$ is the initial density matrix and $e^{Gt}$ is the time-evolution superoperator for the density matrix.

 Calculating the exponential $e^{Gt}$ for a large number of time steps is computationally intensive, but we simplify by selecting a time-step resolution $t_{\text{step}}$ and using an iterative process,
\begin{equation}
 \rho_R(n*t_{\text{step}}) = (e^{G*t_{\text{step}}})^n (\rho(0) \! + \! G^{-1}K) \! - \! G^{-1}K~.
\end{equation}
One exponential is calculated for each selection of parameters in $G$ (including excitation frequency $\omega$), and the problem is reduced to multiple matrix multiplications.
In addition to calculating the matrix exponential, an inverse matrix must also be calculated to solve Eq.~\ref{sol}.
(These two calculations prevent a general analytic solution and consume the most computational time.)
The inverse of the steady-state superoperator $G_S$ needs to be computed once for each selection of parameters, excluding the excitation frequency $\omega$.

\begin{figure*}[th!]
 \includegraphics[width = \textwidth]{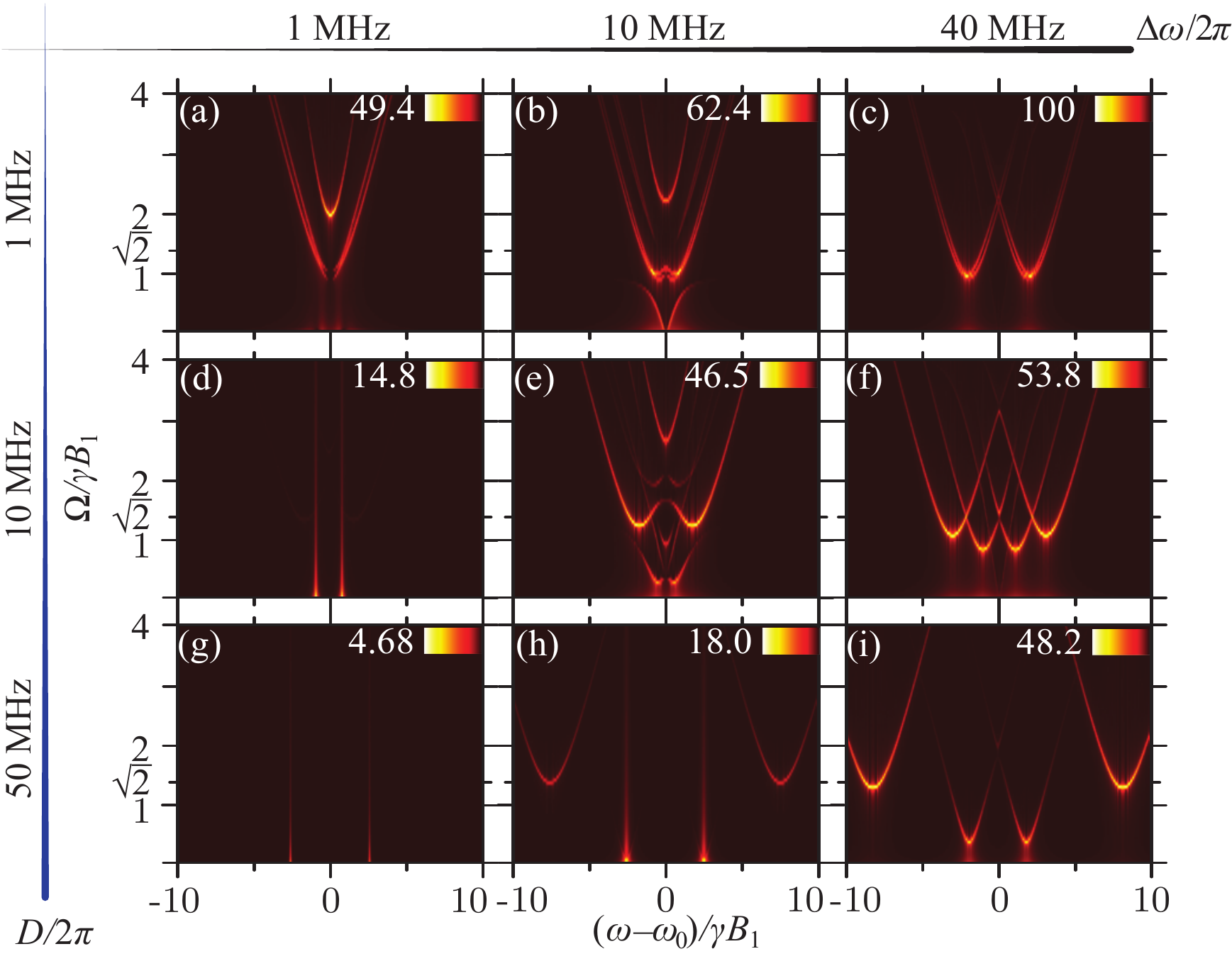}
 \caption{(Color online) Plots of the Fast Fourier Transform FFT$\lbrace Q(\tau) \rbrace$ of the observable $Q(\tau)$ as a function of the excitation frequency $\omega$, in the regime of dipolar coupling only.
The signal intensity for each plot is normalized to plot (c) and given by the number next to the color scale, which indicates the highest magnitude signal intensity in the scale for that plot.
Simulations are done with Larmor separations of $\Delta \omega/2\pi$ = 1 MHz [plots (a), (d), and (g); left column], $\Delta \omega/2\pi$ = 20 MHz [plots(b),(c), and (h); center column], and $\Delta \omega/2\pi$ = 40 MHz [plots (c), (f), and (i); right column]; mapped against dipolar-coupling strengths of $D/2\pi=1$ MHz [plots (a)-(c), first row], $D/2 \pi=10$ MHz [plots (d)-(f), second row], and $D/h=50$ MHz [plots (g)-(i), third row]. The excitation strength is $\gamma B_1 /2 \pi= 10$ MHz.}
 \label{fig:dip_only}
\end{figure*}

Using these techniques we decrease the computation time of $\hat{\rho}(t)$ by three orders of magnitude compared to the conventional ODE solvers that were used in previous studies~\cite{Rajevac_2006,Gliesche_2008,Michel_2009}.
This makes the simulation of complex distributions possible.
For example, the data of Figs.~\ref{fig:J0_distro} and~\ref{fig:J300_distro} are superpositions of 2880 separate simulations generated at a resolution that would be impractical using conventional ODE solvers on a standard personal computer.
We first verified the Liouville-space technique by successfully generating the uncoupled and exchange-coupled simulations previously generated using ODE solvers~\cite{Rajevac_2006,Gliesche_2008}.
Then the simulations obtained for dipolar-coupled pairs (see Figs.~\ref{fig:dip_only} and~\ref{fig:ex_and_dip}) are corroborated by ODE-based simulations (e.g., MATLAB\textsuperscript{\textregistered} solver ODE113).
From our simulations it is possible to describe the nature of the coupling within the pair that leads to experimentally observed spin-dependent transport and recombination processes.

\section{Results and Discussion}
\label{reslut}
The simulations are used to generate a representative database of different coupling strengths and Larmor separations. 
Specifically, we discuss dipolar coupling within the intermediate-spin-pair model to account for $\sqrt{2} \gamma B_1$ Rabi frequencies of experimental pODMR/pEDMR data in disordered semiconductors\cite{Lips_2005,Herring_2009, Lee_2010}.
However, we find that dipolar coupling alone does not account for certain data---exchange coupling must also be included.

For each of the simulations a set of global parameters is used.
Evolving $\hat\rho\left(\tau\right)$ during the application of a 2 $\mu$s excitation pulse, we calculate the observable $Q(\tau)$ with a 4001-step resolution for a range of pulse frequencies $\omega$, where the range of $\omega$ is covered with an 801-step resolution.
We choose $\omega_0/2\pi = 10$ GHz (within the microwave X-band) with the Larmor separation $\Delta \omega$ centered on this value.
For all simulations we also choose a $B_1$ strength such that $\gamma B_1/2\pi = 10$ MHz.
The rate coefficients for singlet recombination, triplet recombination, and dissociation were assigned values of $r_S^{-1} = 1~\mu$s, $r_T^{-1} = 100~\mu$s, and $d^{-1} = 1$ ms, respectively.
These rates represent the characteristic times for spin-dependent recombination or spin-independent dissociation into free charge carriers.
To ensure all singlet information is recorded after the excitation, the observable (Eq.~\ref{observable}) is integrated up to a time $t_0 \! = \!  4r_3^{-1}$, with $r_3$ defined in Eq.~\ref{rates}.
This is done to offset the effects of the inherent signal reduction that arises as the exchange- or dipolar-coupling strength is increased. 
The generation rate $k$ is chosen such that the initial (steady-state) pure-triplet populations of the density matrix are approximately 0.05 ($\rho^S_{11}(0) \! = \! \rho^S_{44}(0) \! \approx \! 0.05$ in the $4 \! \times \! 4$ representation).

All values are taken to be representative of measurement conditions that can be realized in the laboratory, following Ref.~\onlinecite{Gliesche_2008}.
Important physical information garnered from the simulations are the relative positions and amplitudes of the Rabi frequency components $\Omega$ and their dependence on the different coupling strengths.

\subsection{Dipolar Coupling Only}
Here we vary the dipolar-coupling strength $D$ with respect to the Larmor separation $\Delta\omega$, with a negligibly small exchange interaction $J$.
Displayed in Fig.~\ref{fig:dip_only} are simulations with Larmor separations of $\Delta \omega/2\pi = 1$ MHz, $20$ MHz, and $40$ MHz; mapped against dipolar coupling strengths $D/2\pi = 1$ MHz, $10$ MHz, and $50$ MHz.
These are chosen as representative values of $\Delta \omega/2\pi$ and $D/2\pi$ (smaller, approximately equal, and larger) relative to the excitation-field strength $\gamma B_1/2\pi = 10$ MHz.
General features of these data include the resonance curves at $\gamma B_1$, $\sqrt{2}\gamma B_1$, and $2\gamma B_1$, which appear variously as a function of $\Delta \omega/2\pi$ and $D/2\pi$. 
The prominent vertical lines in Fig.~\ref{fig:dip_only}(d) and (g)  result from extremely long integration times compounded with the continuous rotation into ``leaky'' singlet states.

Weak dipolar coupling ($D/2\pi = 1$ MHz) is shown in the top row of Fig.~\ref{fig:dip_only}[(a)-(c)]; we reproduce qualitative features of the weakly coupled pair discussed in Ref.~\onlinecite{Rajevac_2006}.
(The small differences from an uncoupled pair are the result of a slight splitting of the resonances caused by weak dipolar coupling.)
All plots in the top row of Fig.~\ref{fig:dip_only} have on-resonance single-transition Rabi frequencies of $\gamma B_1$.
The multiple-transition Rabi frequencies arise from simple addition and subtraction of the single-transition Rabi frequencies (see Ref.~\onlinecite{Rajevac_2006}).
Thus a weak-dipolar regime leads to no measurable $\sqrt{2}\gamma B_{1}$ Rabi frequency components.
Fig.~\ref{fig:dip_only}(a) has an on-resonance Rabi frequency of 2$\gamma B_{1}$ due to a spin-beating effect from the coherent nutation of both spins~\cite{Rajevac_2006, McCamey_2010}.
Also, the intensity in Fig.~\ref{fig:dip_only}(a) is approximately half that of Fig.~\ref{fig:dip_only}(c); this results directly from the relative triplet/singlet content of the eigenbasis in Eq.~\ref{basis}.
The middle row of Fig.~\ref{fig:dip_only} has an intermediate-dipolar strength ($D/2\pi = 10$ MHz $=\gamma B_1$) and no strong $\sqrt{2} \gamma B_1$ Rabi frequency components.
Indeed this $\sqrt{2} \gamma B_1$ component is barely visible in Fig.~\ref{fig:dip_only}(d), much weaker than the bright vertical lines.   

Strong dipolar coupling ($D/2\pi = 50$ MHz) is shown in the last row of Fig.~\ref{fig:dip_only}.
When the Larmor separation is less than the dipolar strength, $\Delta \omega \! < \! \gamma B_1 \! < \! D$ [Fig.~\ref{fig:dip_only}(g)], there is a weak non-visible (due to bin size) transition with a $\sqrt{2} \gamma B_1$ Rabi frequency.
Both Fig.~\ref{fig:dip_only}(h) and Fig.~\ref{fig:dip_only}(i) show a strong $\sqrt{2} \gamma B_1$ Rabi frequency component.
Fig.~\ref{fig:dip_only}(h) has a dipolar-coupling strength greater than the Larmor separation, and both are greater than or comparable to the excitation strength, $D \! > \!( \Delta \omega \! \approx \! \gamma B_1$).
Fig.~\ref{fig:dip_only}(i) has a Larmor frequency separation and dipolar-coupling strength approximately equal, but both greater than the excitation strength, ($\Delta \omega \! \approx \! D) \! > \! \gamma B_1$.
Thus a Rabi frequency of $\sqrt{2} \gamma B_1$ only occurs in the regime where dipolar-coupling strength is greater than both the Larmor separation and the excitation strength, $D \geq \Delta \omega , D > \gamma B_1$.
The limits of this regime are discussed in Sect.~\ref{dipole}.

Each column of Fig.~\ref{fig:dip_only} reflects the observable intensity getting weaker with increasing dipolar-coupling strength; this is because we are approaching a triplet-singlet energy eigenbasis.
Another trend occurring down each column is the separation of the on-resonance positions increasing with dipolar-coupling strength, also demonstrated with Eq.~\ref{strD}.

\begin{figure}
\includegraphics{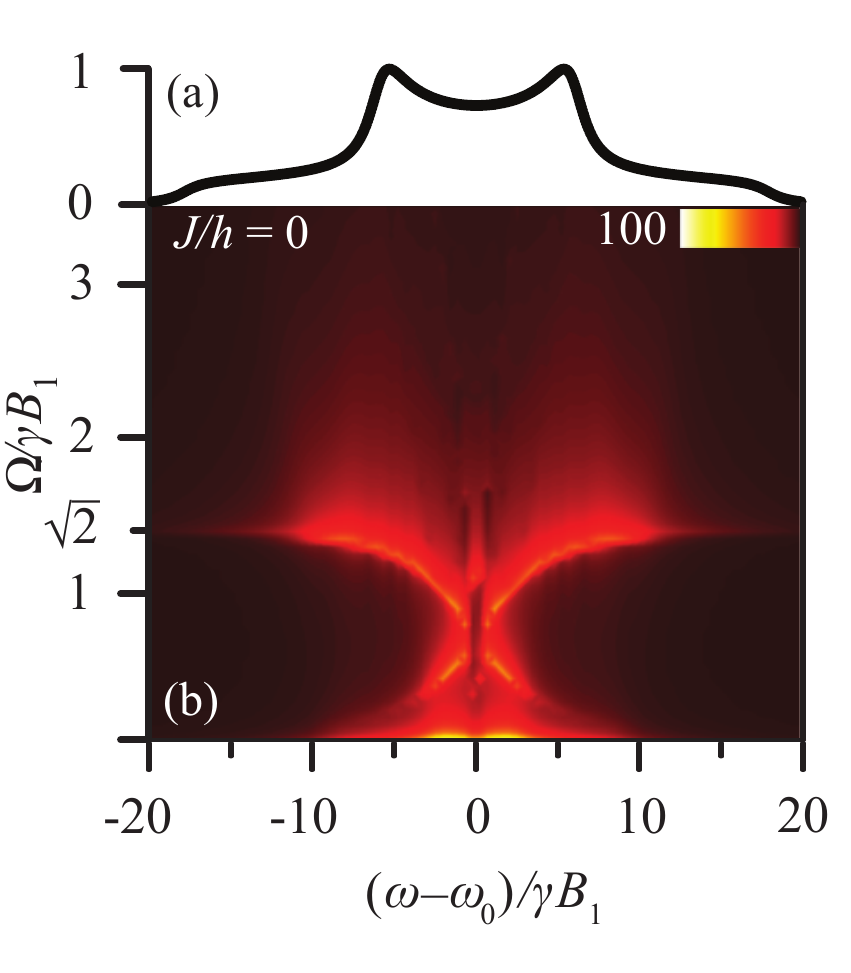}
 \caption{ \label{fig:J0_distro}(Color online) (a) The distribution of dipolar-coupling strengths for the simulation shown in (b). The distribution is a Pake doublet with Larmor separation $\Delta \omega/2\pi = 40$ MHz and dipolar-coupling strength of $D/2\pi = 80$ MHz convoluted with a Lorentzian with a half-width of $10$ MHz.
(b) Plot of the Fast Fourier Transform FFT$\lbrace Q(\tau) \rbrace$ of the observable $Q(\tau)$ as a function of the excitation frequency $\omega$.
The signal intensity is normalized and given by the number next to the color scale, which indicates the highest magnitude signal intensity.
The simulation uses the distribution in Fig.~\ref{fig:J0_distro}(a), with no exchange coupling $J = 0$.
The excitation strength is $\gamma B_1/2\pi = 10$ MHz.}
\end{figure}
\begin{figure*}[htbp]
\includegraphics[width = \textwidth]{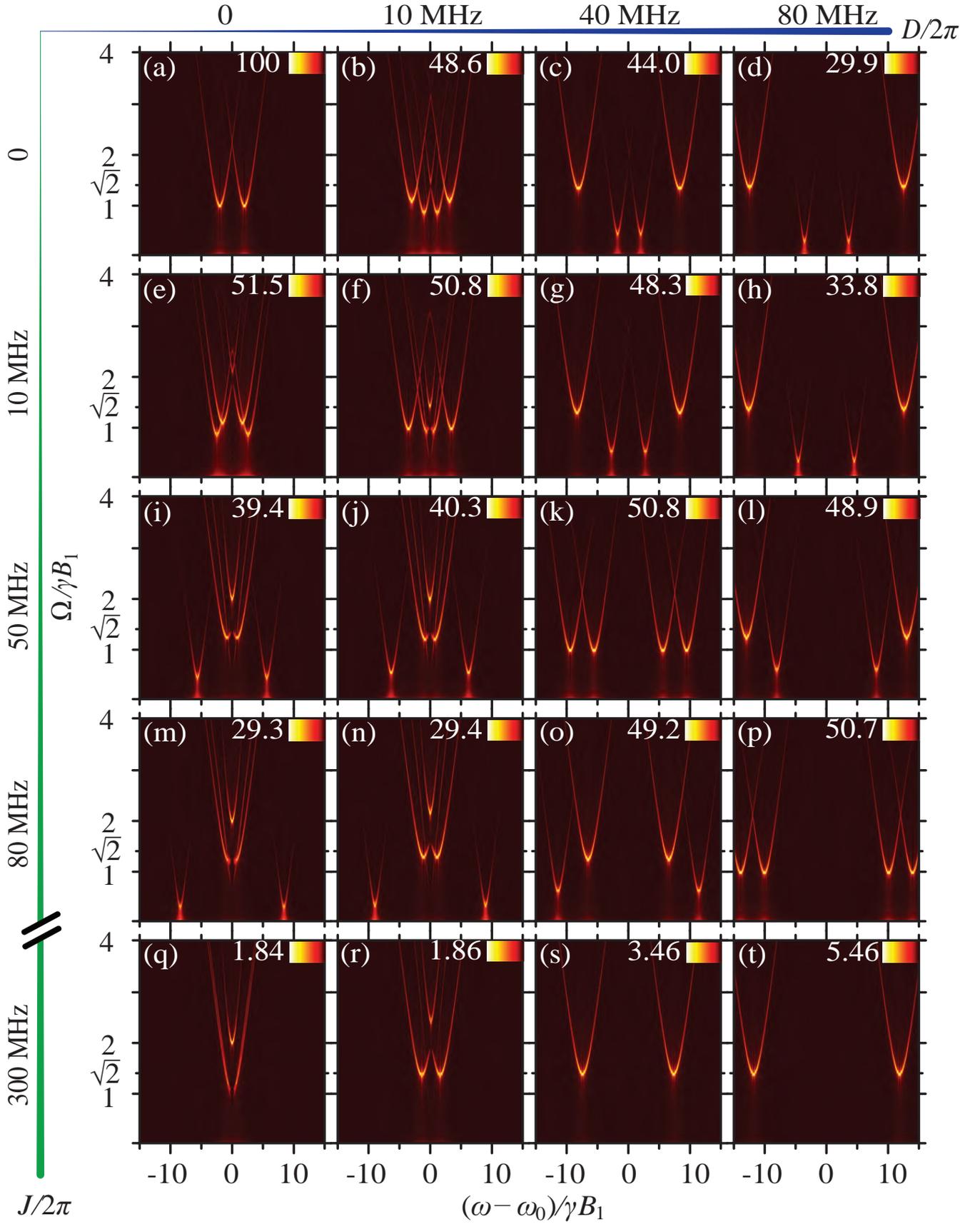}
 \caption{ \label{fig:ex_and_dip}(Color online) See Sect.~\ref{dip_Ex} for description.}
\end{figure*}

We turn now to modeling a more realistic pEDMR/pODMR signal.
We operate in a regime of strong $\sqrt{2} \gamma B_1$ Rabi components and seek to model an actual spin-pair distribution in a disordered material. 
Indeed, many materials with pronounced spin-selection rules are disordered semiconductors, including those for which the significance of the dipolar interaction has been discussed~\cite{Lips_2005,Herring_2009,Lee_2010}.
In a disordered environment, the orientation of a spin pair with respect to an applied magnetic field can be entirely random.
The strengths of the dipolar fields are highly orientation dependent because of the inherently anisotropic spin-dipolar interaction, even if a fixed spin-pair distance is considered (rather than a distribution of distances).
The well-known Pake distribution accounts for this random orientation~\cite{Pake_1948}.
Fig.~\ref{fig:J0_distro} is a simulation using a Pake distribution with a dipolar coupling strength of $D/2\pi = 80$ MHz, a Larmor separation of $\Delta \omega/2\pi = 40$ MHz, and $J = 0$.
Fig.~\ref{fig:J0_distro}(a) is created using a 2880-point Pake distribution convolved with a 10 MHz FWHM Lorentzian function to account for power broadening due to the excitation pulse.
We then generate simulations for the 2880 dipolar-coupling strengths and, using relative weights from Fig.~\ref{fig:J0_distro}(a), average those simulations to make Fig.~\ref{fig:J0_distro}(b).

A comparison of Fig.~\ref{fig:J0_distro}(b) with experimental pEDMR and pODMR data~\cite{Lips_2005,Herring_2009,Lee_2010} strongly supports the notion that the strong transitions with a Rabi frequency of $\sqrt{2} \gamma B_1$ arise from a strong dipolar interaction.
However, other characteristics of Fig.~\ref{fig:J0_distro}(b) do not match experimental data.
The strong low-Rabi-frequency components ($\approx 0.4 \gamma B_1$) of Fig.~\ref{fig:J0_distro}(b) are not seen in pODMR of hydrogenated amorphous silicon (a-Si:H), as seen in Fig.~8 of Ref.~\onlinecite{Lips_2005} or Fig.~1a of Ref.~\onlinecite{Herring_2009}.
Low-Rabi-frequency components ($\approx 0.1$-$0.2~\gamma B_1$) are seen in pEDMR of hydrogenated amorphous silicon nitride (a-SiN$_x$:H), given in Fig.~2(e) of Ref.~\onlinecite{Lee_2010}.
However, the same data also shows a $\sqrt{2} \gamma B_1$ Rabi frequency relatively flat with respect to excitation frequency compared to the curved shape in Fig.~\ref{fig:J0_distro}(b).
From these discrepancies we conclude that dipolar coupling alone cannot account for the pODMR/pEDMR data reported in the literature.

\subsection{Dipolar and Exchange Coupling}

\label{dip_Ex}
We now introduce exchange coupling between the spin pairs in addition to the dipolar coupling. The parameter space for the simulation is quickly growing; we give a small representation in Fig.~\ref{fig:ex_and_dip}.

Fig.~\ref{fig:ex_and_dip} has plots of the Fast Fourier Transform FFT$\lbrace Q(\tau) \rbrace$ of the observable $Q(\tau)$ as a function of the excitation frequency $\omega$.
The signal intensity for each plot is normalized to plot (a) and given by the number next to the color scale, which indicates the highest magnitude signal intensity in the scale for that plot.
Simulations are done with a Larmor separation of $\Delta \omega/2\pi \!= \!40$ MHz.
Dipolar-coupling strengths are $D/2\pi\!=\!0$ [plots Fig.~\ref{fig:ex_and_dip}(a),(e),(i),(m),(q), first column], $D/2\pi\!=\!10$ MHz [plots Fig.~\ref{fig:ex_and_dip}(b),(f),(j),(n),(r), second column], $D/2\pi\!=\!40$ MHz [plots Fig.~\ref{fig:ex_and_dip}(c),(g),(k),(o),(s), third column], $D/2\pi\!=\!80$ MHz [plots Fig.~\ref{fig:ex_and_dip}(d),(h),(l),(p),(t), fourth column]; mapped against exchange-coupling strengths of $J/2\pi\!=\!0$ [plots Fig.~\ref{fig:ex_and_dip}(a)-(d), first row], $J/2\pi\!=\!10$ MHz [plots Fig.~\ref{fig:ex_and_dip}(e)-(h), second row], $J/2\pi\!=\!50$ MHz [plots Fig.~\ref{fig:ex_and_dip}(i)-(l), third row], $J/2\!\pi=\!80$ MHz [plots Fig.~\ref{fig:ex_and_dip}(m)-(p), fourth row], $J/2\pi \!= \!300$ MHz [plots Fig.~\ref{fig:ex_and_dip}(q)-(t), fifth row].
With the expection of Fig.~\ref{fig:ex_and_dip}(a), dipolar- and exchange-coupling strengths are chosen greater than or equal to  the excitation strength $\gamma B_1/2 \pi\! =\! 10$ MHz.

Fig.~\ref{fig:ex_and_dip}(a) is an uncoupled spin pair that satisfies the weak-coupling limit described in Sect.~\ref{weak}; two resonances are located at the Larmor frequencies of the electron and hole and have Rabi frequencies of $\gamma B_1$.
The uncoupled spin pair yields the maximum relative intensity (100) in Fig.~\ref{fig:ex_and_dip}.
Fig.~\ref{fig:ex_and_dip}(f) and Fig.~\ref{fig:ex_and_dip}(p) are in the effectively-weak-coupling limit also described in Sect.~\ref{weak}, where the dipolar- and exchange-coupling strengths are equal ($J \! = \! D$).
Fig.~\ref{fig:ex_and_dip}(k) has approximately equal dipolar- and exchange-coupling strengths with on-resonance Rabi frequencies slighty offset from $\gamma B_1$.

The first row of Fig.~\ref{fig:ex_and_dip} is similar to the third column of Fig.~\ref{fig:dip_only}; there is no exchange interaction present and the relative intensity of FFT$\lbrace Q(\tau) \rbrace$ decreases with increasing dipolar coupling strength.
The distribution in Fig.~\ref{fig:J0_distro}(b) can be thought of as generated from intermediate values between and including Fig.~\ref{fig:ex_and_dip}[(a)-(d)].
The sequence across the first row of Fig.~\ref{fig:ex_and_dip} best illustrates the discussion in Sect.~\ref{dipole}; the two $\ket{T} \! \leftrightarrow \! \ket{2}$ transitions are split from the center frequency $\omega_0$ and trend upwards to the strong dipolar-coupling limit with a $\sqrt{2} \gamma B_1$ Rabi frequency.
The two $\ket{T} \! \leftrightarrow \! \ket{3}$ transitions are also split from the center frequency but are approaching their strong-coupling limit (zero Rabi frequency).

As discussed in the strong-exchange-coupling limit of Sect.~\ref{exchange}, the single-transition analysis fails to account for the observed Rabi frequencies; this is explicitly seen down the first column of Fig.~\ref{fig:ex_and_dip}, for the two $\ket{T} \! \leftrightarrow \! \ket{3}$ transitions.
Indeed the observed frequencies are not $\sqrt{2} \gamma B_1$ and 0 as would be obtained from Eq.~\ref{Rabio}; multiple transitions must be considered to obtain the correct values.
The first column in Fig.~\ref{fig:ex_and_dip} is an extension of the simulations shown in the third column of Fig.~2 in Ref.~\onlinecite{Gliesche_2008}, where the exchange interaction is considered without the dipolar interaction.
With increasing exchange-coupling strength the $\ket{T} \! \leftrightarrow \! \ket{2}$ transitions are split further about $\omega_0$, while the $\ket{T} \! \leftrightarrow \! \ket{3}$ on-resonance frequency positions remain unaffected.
This is seen down each column of Fig.~\ref{fig:ex_and_dip} and from Sects.~\ref{exchange} and \ref{both}.
The role dipolar coupling plays is shown by a common trend throughout all rows of Fig.~\ref{fig:ex_and_dip}.
An increase of dipolar-coupling strength creates a greater energy splitting, causing a particular transition to tend further from $\omega_0$ and making the single-transition analysis of Eq.~\ref{Rabio} valid.
Therefore, a $\sqrt{2} \gamma B_1$ Rabi frequency is present if the dipolar coupling is strong enough.
From these general trends, we determine that \emph{only} the combination of strong dipolar and even stronger exchange (see Fig.~\ref{fig:ex_and_dip}[(r)-(t)]) yields strong $\sqrt{2} \gamma B_1$ Rabi frequency components without any strong low-frequency (0-$\gamma B_1$) components.

\begin{figure}[rt!]
\includegraphics{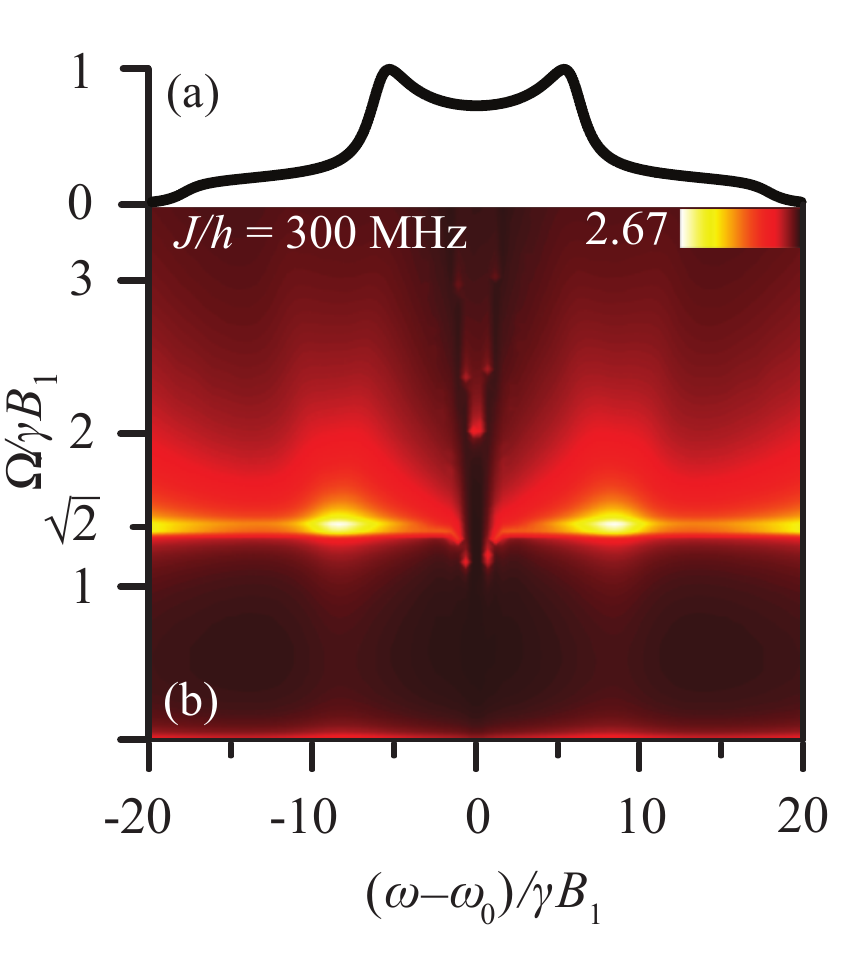}
\caption{ \label{fig:J300_distro}(Color online) (a) The distribution of dipolar-coupling strengths for the simulation shown in (b). The distribution is a Pake doublet with Larmor separation $\Delta \omega/2\pi = 40$ MHz and dipolar-coupling strength of $D/2\pi = 80$ MHz convoluted with a Lorentzian with a half-width of $10$ MHz.
(b) Plot of the Fast Fourier Transform FFT$\lbrace Q(\tau) \rbrace$ of the observable $Q(\tau)$ as a function of the excitation frequency $\omega$.
The signal intensity is normalized to Fig.~\ref{fig:J0_distro}(b) and given by the number next to the color scale, which indicates the highest magnitude signal intensity.
The simulation uses the distribution in Fig.~\ref{fig:J300_distro}(a), with an exchange-coupling strength of $J/2\pi = 300$ MHz. The excitation strength is $\gamma B_1/2\pi = 10$ MHz.}
\end{figure}
Using this analysis we generate Fig.~\ref{fig:J300_distro}(b), which shows a distribution simulation similar to that of Fig.~\ref{fig:J0_distro}(b) but with a strong exchange coupling.
This distribution samples from the regime where there is a large difference between exchange- and dipolar-coupling strengths with $J \! > \! D \! > \! \Delta \omega \! > \! \gamma B1,~J-D \! \gg \! \Delta \omega$.
Fig.~\ref{fig:J300_distro}(b) has a flat Rabi frequency of $\sqrt{2} \gamma B_1$ and exhibits no strong low-Rabi-frequency components.
It also exhibits some $2 \gamma B_1$ components.
These same characteristics are found in the experimental data of Ref.~\onlinecite{Lips_2005,Herring_2009, Lee_2010}.

The pODMR data of a-Si:H in Fig.~8 of Ref.~\onlinecite{Lips_2005} is almost identical to Fig.~\ref{fig:J300_distro}(b), showing strong $\sqrt{2} \gamma B_1$ Rabi-frequency components, weak components around $2 \gamma B_1$, and no low-frequency components.
Thus our spin-pair model predicts that both dipolar and exchange coupling are responsible for the pODMR data of Ref.~\onlinecite{Lips_2005}.
Moreover, the simulations show that the relative coupling strengths present in this data are in a regime with a large difference in exchange- and dipolar-coupling strengths, with $J\!> \!D$.
This analysis supports the discussion presented in Ref.~\onlinecite{Lips_2005} that suggested dipolar coupling was the cause for the observed data; it further predicts that strong exchange coupling was also present.

The pODMR of a-Si:H geminate pairs in Fig.~3a of Ref.~\onlinecite{Herring_2009} is also very similar to the simulation in Fig.~\ref{fig:J300_distro}(b), with the caveat that there appears to be the presence of uncoupled spins that produce strong $\gamma B_1$ Rabi frequencies.
In that data set, the strong transitions with a Rabi frequency of $\sqrt{2} \gamma B_1$ are flat with respect to excitation frequency but become abruptly weaker; this is characteristic of the Pake distribution in Fig.~\ref{fig:J300_distro}(b) which also has a strong $\sqrt{2} \gamma B_1$ component becoming abruptly weaker at an excitation frequency of $(\omega - \omega_0)/\gamma B_1\! = \! 10$.
This experimental data also has no lower-Rabi-frequency components (0-$\gamma B_1$), which we have shown to be a defining characteristic of the regime in which there is a large difference in exchange- and dipolar-coupling strengths with $J \! > \! D$.
Therefore, we determine that the geminate pairs show the characteristics of uncoupled pairs mixed with strongly dipolar-coupled pairs discussed in Ref.~\onlinecite{Herring_2009}, but we also predict the presence of a strong exchange coupling.

Finally, the pEDMR data in Fig.~3e of Ref.~\onlinecite{Lee_2010} show broad $\sqrt{2} \gamma B_1$ Rabi frequency components and weak $\gamma B_1$ Rabi frequency components.
Again, this could be characteristic of a resonance involving mostly uncoupled pairs and some strong exchange- and dipolar-coupled pairs.
However, the presence of both strong dipolar and exchange coupling cannot explain the strong low-frequency ($\approx 0.2 \gamma B_1$) components present in the Ref.~\onlinecite{Lee_2010} data.
Perhaps the curvature leading to the $\sqrt{2} \gamma B_1$ limit seen in Fig.~\ref{fig:J0_distro} cannot be seen in the Ref.~\onlinecite{Lee_2010} data because of a low number of dipolar-coupled pairs.
However, if the strong low-frequency components are due to strong dipolar coupling alone, we would expect (from Fig.~\ref{fig:J0_distro}) that the $\sqrt{2} \gamma B_1$ component would be as strong as the low-frequency component, and the data does not have this feature. 
Therefore we conclude that dipolar-coupled pairs can explain the $\sqrt{2} \gamma B_1$ Rabi frequencies in the a-SiN$_x$:H data presented in Ref.~\onlinecite{Lee_2010}, but whether exchange is present cannot be confirmed or rejected due to the weak signal strength of the strongly coupled pairs relative to the uncoupled pairs in that data.

\section{Summary and Conclusion}
Numerical and analytical methods were used to investigate the role of the dipolar interaction for electrically and optically detected Rabi oscillation frequencies of intermediate-spin-pair systems.
A general description of the physics of pEDMR and pODMR transient-nutation experiments was given that includes dipolar and exchange interactions, the Larmor separations within the intermediate pairs, and the excitation-field strength.
We have presented a intermediate-spin-pair model that corroborates previous numerical studies that included weakly coupled pairs only~\cite{McCamey_2010,Rajevac_2006} and exchange-coupled pairs only~\cite{Gliesche_2008}.
The model also supports experimental studies that attributed the observation of $\sqrt{2} \gamma B_1$ Rabi-frequency components with pODMR/pEDMR of disordered semiconductors~\cite{Lips_2005,Herring_2009, Lee_2010} to the presence of strong dipolar coupling within the spin-pair model.
We have shown that pODMR data of a-Si:H presented in Ref.~\onlinecite{Lips_2005,Herring_2009} can be explained within a intermediate-pair model in the regime of strong dipolar coupling and stronger exchange coupling, $(J-D)^2 \gg \Delta \omega^2,~J \! > \! D \! > \! \Delta \omega \! > \! \gamma B_1$. 

\begin{acknowledgments}
The authors thank A.~D.~Ballard and K.~van Schooten for helpful discussions.
We acknowledge the support of this work by the National Science Foundation through the Materials Research Science and Excellence Center (\#DMR-1121252).
CB also acknowledges support through the National Science Foundation CAREER award (\#0953225).
\end{acknowledgments}

\end{document}